\documentclass[sigconf]{acmart}
\usepackage{pgfplots}
\usepackage{subfigure}
\usepackage{tabularx}
\usepackage{booktabs}
\usepackage{multirow}
\usepackage{pgfplots}
\pgfplotsset{compat=1.18}
\usepgfplotslibrary{groupplots}

\usepackage{caption}
\usetikzlibrary{patterns} 

\AtBeginDocument{%
  }

\setcopyright{acmlicensed}
\copyrightyear{2018}
\acmYear{2018}
\acmDOI{XXXXXXX.XXXXXXX}
\acmConference[Conference acronym 'XX]{Make sure to enter the correct
  conference title from your rights confirmation email}{June 03--05,
  2018}{Woodstock, NY}
\acmISBN{978-1-4503-XXXX-X/2018/06}

\begin{document}

\title{Membership Inference Attacks on In-Context Examples in LLM-based Recommender Systems}

\author{Jiajie He}
\email{jiajieh1@umbc.edu}
\affiliation{%
  \institution{University of Maryland, Baltimore County}
  \city{Baltimore}
  \state{MD}
  \country{USA}
}

\author{Min-chen Chen}
\email{mchen12@umbc.edu}
\affiliation{%
  \institution{University of Maryland, Baltimore County}
  \city{Baltimore}
  \state{MD}
  \country{USA}
}

\author{Xintong Chen}
\email{chen3xt@mail.uc.edu}
\affiliation{%
  \institution{University of Cincinnati}
  \city{Cincinnati}
  \state{OH}
  \country{USA}
}

\author{Xinyang Fang}
\email{xinyangf@usc.edu}
\affiliation{%
  \institution{University of Southern California}
  \city{Los Angeles}
  \state{CA}
  \country{USA}
}

\author{Yuechun Gu}
\email{ygu2@umbc.edu}
\affiliation{%
  \institution{University of Maryland, Baltimore County}
  \city{Baltimore}
  \state{MD}
  \country{USA}
}
\author{Keke Chen}
\email{kekechen@umbc.edu}
\affiliation{%
  \institution{University of Maryland, Baltimore County}
  \city{Baltimore}
  \state{MD}
  \country{USA}
}

\renewcommand{\shortauthors}{Trovato et al.}

\begin{abstract}
Large language models (LLMs) based recommender systems (RecSys) can adapt flexibly across different domains. It uses in-context learning (ICL), i.e., prompts, including sensitive historical user-specific item interactions, to customize the recommendation functions. However, no study has examined whether such private information may be exposed by novel privacy attacks. We design two membership inference attacks (MIAs): \emph{ItemMem}, and \emph{RecInertia}, aiming to identify whether system prompts contain the victim's information. We have carefully evaluated them on the latest open-source LLMs and three well-known RecSys datasets. The results confirm that the MIA threat to LLM RecSys is realistic and can be more sophisticated than prompt extraction. They utilize the unique prompt structures in ICL RecSys and cannot be easily mitigated with existing defense methods on prompt extraction. 
\end{abstract}


\keywords{Do, Not, Use, This, Code, Put, the, Correct, Terms, for,
  Your, Paper}

\received{20 February 2007}
\received[revised]{12 March 2009}
\received[accepted]{5 June 2009}

\maketitle

\section{Introduction}
\label{sec:introduction}
Recommendation systems (RecSys) have seen significant advances over the past decade and are widely used across scenarios such as job matching, e-commerce, and entertainment. However, one critical challenge remains: recommendation models are naturally task-specific, as they are typically trained on task-specific user-item interactions \cite{liu2023chatgptgoodrecommenderpreliminary,Zhao_2024,zhang26,zang22}. Other significant challenges also remain, such as the user cold start problem \cite{zhang19,yuan23} and semantic understanding of user intentions \cite{zhang19,jannach23}. Practitioners and researchers have been looking for more efficient approaches to addressing these challenges. 

The use of in-context learning (ICL) with large language models (LLMs) offers capabilities that traditional recommender systems struggle to achieve, particularly in settings requiring flexibility, rapid adaptation, and rich semantic understanding. Unlike conventional collaborative filtering or deep learning–based recommenders that rely on extensive training and fixed model parameters, ICL-based recommender systems can incorporate new user information, preferences, and contextual signals directly at inference time without retraining. This enables effective handling of cold-start users, dynamic user intent, and cross-domain recommendations \cite{liu2023chatgptgoodrecommenderpreliminary,He_2023,hou2024large,dai2023uncovering, wang-lim-2024-whole}. Empirical results show that ICL-based RecSys can now achieve comparable performance in some tasks compared to traditional RecSys~\cite{wang-lim-2024-whole}. 
Moreover, LLMs can leverage unstructured data—such as natural language descriptions, reviews, and conversational context—allowing recommendations \cite{He_2023,jannach23} to be more interpretable and aligned with user intent. These advantages make ICL-based recommender systems especially attractive for emerging applications such as conversational recommendation and personalized assistants. Motivated by these advantages, industrial practitioners, such as Amazon \cite{Liang2025} and Google \cite{52472}, have also started incorporating ICL-based LLM RecSys in production. It has become an important direction for next-generation RecSys~\cite{li2024large,wu2024surveylargelanguagemodels}. 

At the same time, this paradigm introduces new design considerations, as user data is directly embedded in prompts, not in training time, motivating a closer examination of both its benefits and potential risks. Prior work has focused on improved prompting strategies, such as aggregated demonstrations \cite {wang-lim-2024-whole} or optimized example selection \cite{zhang-etal-2025-llmtreerec} for stable performance. However, the security and privacy issues are largely unexplored, in particular, privacy leakage through prompts. 

\textbf{Unique features of ICL RecSys MIAs.} One of the most fundamental privacy attacks is the membership inference attack (MIA) \cite{hu2022membership} that tries to determine whether a record is used in the model's training dataset. While recent studies on RecSys MIAs focus on traditional RecSys models \cite{zhang2021membership,yuan2023interaction,zhong2024interaction,10.1145/3705328.3748052}, there is no study on ICL-based RecSys, which has several unique features that existing methods cannot address.

(1) Traditional RecSys MIAs target the training data encoded in model parameters, while
ICL RecSys MIAs target data explicitly injected at inference time via prompts. Since ICL RecSys does not depend on large-scale users' interaction patterns like the training data used by collaborative filtering, MIA methods working on traditional RecSys, e.g., the similarity-based MIA \cite{zhang2021membership, yuan2023interaction, zhong2024interaction, 10.1145/3705328.3748052, zhu2023membership},  do not work for ICL RecSys (as our experiments show).

(2) Existing RecSys MIAs assume that the adversary knows the training data distribution. It is used to generate training data for offline shadow models that mimic the behavior of the target model. In LLM-based RecSys, only a few training examples appear in system prompts. The concept of shadow models requires re-examination. 

(3) ICL RecSys utilizes some distinct LLM features, such as memorization \cite{carlini23memorization} and reasoning \cite{open-ai-reasoning}, that the traditional RecSys does not have. These features might enable new attacks \cite{wen2024membershipinferenceattacksincontext} that are distinct from those on traditional RecSys models. 

\textbf{Prompt extraction vs MIA.}
Recent studies investigate prompt extraction attacks, where adversaries attempt to recover hidden prompts or demonstrations through model queries \cite{perez22,carlini23memorization,sha24}. Once they are successful, they can be directly applied for MIA. However, follow-up defenses, such as prompt sanitization and output filtering \cite{edemacu25}, can significantly mitigate direct extraction. We notice the significant difference between MIAs and prompt extraction. Prompt extraction treats the prompt as a static secret to be reconstructed, without exploiting the task-specific structure of the underlying problem. As a result, such attacks do not fully capture the privacy risks in structured applications like recommendation, where demonstrations encode rich user interaction patterns rather than isolated text snippets. The question is how to design such MIAs in ICL RecSys that cannot be defended by existing prompt-extraction defense methods.

A systematic understanding of such emerging MIA threats is crucial, as it enables designers of LLM-based RecSys to identify potential privacy vulnerabilities and proactively integrate appropriate privacy protection mechanisms into system design. 

\textbf{Scope of Our Research.} We design, evaluate, and analyze two membership inference attacks on LLM-powered RecSys that uses in-context learning to customize its recommendation function. These attacks target private user-item interactions 
embedded in system prompts by the LLM RecSys provider. We follow the previous black-box setting \cite{zhang2021membership} and assume that the attacker knows the victim user's historical interactions and sample recommendations, but is unaware of whether the LLM RecSys utilized any of such data to compose the system prompts.

These attacks include (1) \textbf{Item Memorization (ItemMem)} attacks, which exploit the inherent memorization capability of LLMs, and (2) \textbf{Recommendation Inertia (RecRecInertia)} attacks, which use perturbed user queries to indirectly infer membership information.

We conducted extensive experiments on six popular open-source large language models (Llama3:8b, Llama4:109b Gemma3:4b, Mistral:7b, GPT-OSS:20b and GPT-OSS:120b) and three classical benchmark datasets: MovieLens-1M, Amazon Book, and Amazon Beauty Products. Empirical results show that ItemMem and RecInertia are quite effective. 
We also investigated whether well-known defense methods: prompt sanitization and output filtering, work on these attacks. While both can reduce the effectiveness of attacks in a few cases, the attacks remain effective overall. We further analyze factors influencing successful attacks, including the number of shots used by system prompts, the positions of the attacked shots in the prompt, and the number of poisoned items in RecInertia.

Our contributions can be summarized as follows: 
\begin{itemize}
\item To the best of our knowledge, we are the first to propose and study membership inference attacks against ICL-powered LLM RecSys, which utilize the unique structural features in the ICL demonstrations for RecSys.
\item We have designed two effective MIAs: ItemMem and RecInertia attacks, and shown that the existing prompt-extraction defense methods do not work satisfactorily against them.   
\item We have conducted extensive experiments to evaluate the performance of these attacks and examined the factors that influence their performance. 
\end{itemize}

\section{System Architecture and Threat Model}

\subsection{ICL RecSys} 
ICL has proven effective in adapting LLMs to various downstream tasks, particularly in recommendation systems (RecSys).  Its success stems from the design of prompts and in-context demonstrations \cite{gao2021makingpretrainedlanguagemodels,liu2023chatgptgoodrecommenderpreliminary,Zhao_2024,zhiyuli2023bookgptgeneralframeworkbook}. Role-based textual descriptions in the prompt, such as ``You are a book rating expert...'' can also improve the performance \cite{zhiyuli2023bookgptgeneralframeworkbook}. 

\begin{figure}[htbp]
  \centering
  \vspace{-3mm}
  \includegraphics[width=0.7\linewidth]{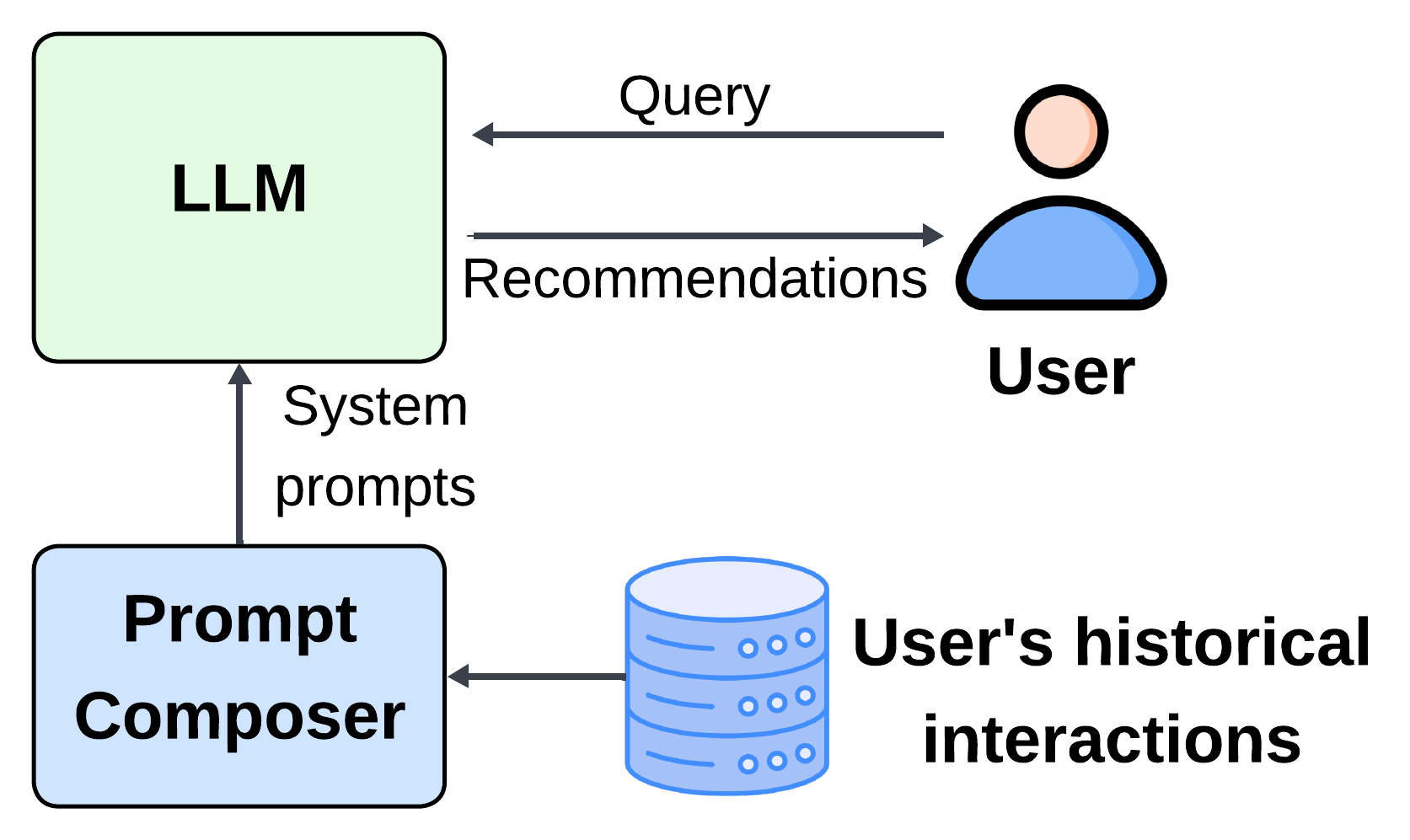}
  \caption{System Architecture for ICL-RecSys}
  \vspace{-2mm}
  \label{fig:ICL-RecSys}
\end{figure}

Figure \ref{fig:ICL-RecSys} shows a typical ICL-based LLM RecSys architecture. The core components include the LLM, a database containing the historical interactions of multiple users, and a prompt composer that can adapt to the recommendation task and each user's specific preferences. Demonstrations are retrieved from the service provider's internal database using a specific strategy, e.g., matching the user's profile. A typical query looks like
``The user has watched the following movies: x, y, z. Based on this watch history, please recommend the top 10 movies in descending order that the user is most likely to watch next. Format the output as a numbered list of movie titles only...'', which is composed with the system prompt and the demonstrations to form the final prompt. 

\subsection{Threat Model}
\label{sec:threat model}
\textbf{Adversary's Objective.}
The primary objective of the adversary is to determine whether a specific target user $u$, we call \textbf{the victim}, was included in the construction of a prompt used to customize a language model $\mathcal{M}$. In the RecSys setting, we also have a set of items, $I$, for which each user may interact selectively based on their preferences. The prompt in the ICL RecSys, denoted as \textit{p}, comprises a set of $k$ demonstrations, formatted as:

\emph{p} = \{\emph{Task Instruction}, \emph{Recommendation Examples:} \\
$(u_1, I_1)\rightarrow R_1,  \dots, (u_k, I_k)\rightarrow R_k \}$

where $u_i$ are from the user set $U$, $I_i$ is $u_i$'s interaction set, $I_i \in I$, and $R_i$ is the set of recommended items, $R_i \in I$. The adversary's goal is to determine whether the victim $u$ appears in the system prompt $u \overset{?}{\in} \{u_1, \dots, u_k\}$. 

\textbf{Adversary's Capabilities.}
The adversary can access the LLM used in the RecSys, the victim's historical interactions and recommendations. We consider the most strict and realistic scenario, where the adversary has only black-box access to the target language model $\mathcal{M}$ and its recommended items, but not the tokenizer or the associated output token probabilities.  We also assume the adversary can access general-purpose word embeddings generated by open-source LLMs (not the target LLM), which can be used, for example, to launch some attacks discovered in traditional RecSys \cite{zhang2021membership,yuan2023interaction,zhong2024interaction,10.1145/3705328.3748052}.

\section{Membership Inference Attack Methods on ICL RecSys}\label{sec:attack methods}
In the following, we present two membership inference attacks on ICL-based RecSys that explore the unique features of ICL RecSys: the item memorization and recommendation inertia attacks. 


\subsection{Item Memorization Attack}
Simple prompt extraction attacks, such as ``have you seen the user interact with x, y, z? '' or ``repeat the recommendation examples, can also extract the private information.'' However, they are quite easy to defend with input sanitization methods \cite{edemacu25}, as we will show in experiments. To make the attack more disguised, we want to explore the unique features in ICL RecSys. 

\begin{figure}[htbp]
  \centering
  \vspace{-4mm}
  \includegraphics[width=0.7\linewidth]{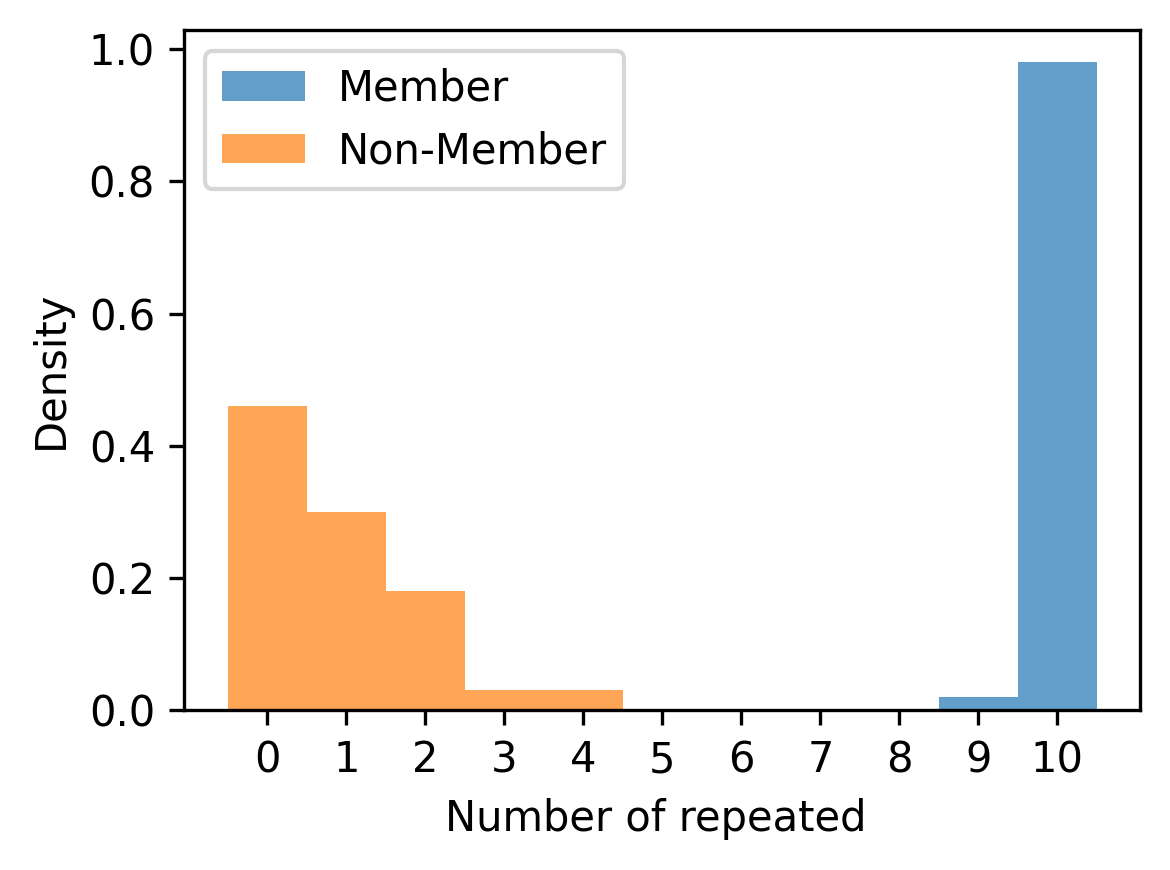} 
  \caption{The repeated items between members and non-members.}
  \vspace{-4mm}
  \label{fig:semantic_hist}
\end{figure}

Specifically, we should design attacks that look normal to queries in ICL RecSys. We experimented with standard queries using ICL RecSys and observed an important feature: if the LLM has seen the victim's demonstration in the prompt, it is more likely to include the recommended items from the demonstration in the output. Figure \ref{fig:semantic_hist} shows the distinct distribution between the repeated items between members and non-members. It inspired us to design the item memorization (ItemMem) attack. This attack uses normal recommendation requests, which cannot be blocked via prompt sanitization. 

\textbf{Hypothesis.} This attack leverages the pattern memorization capability of language models to detect context-aware responses. When provided with the victim user $u$ and the user's interacted item set $I_u$ and recommended item $R^h_u$ in a demonstration case, we hypothesize that the model may learn the recommendation pattern, $I_u \rightarrow R^h_u$, and attempt to repeat the recommended items in the output. 

\noindent \textbf{Method.} It consists of the following steps:

\begin{figure}[htbp]
  \centering
  \includegraphics[width=0.8\linewidth]{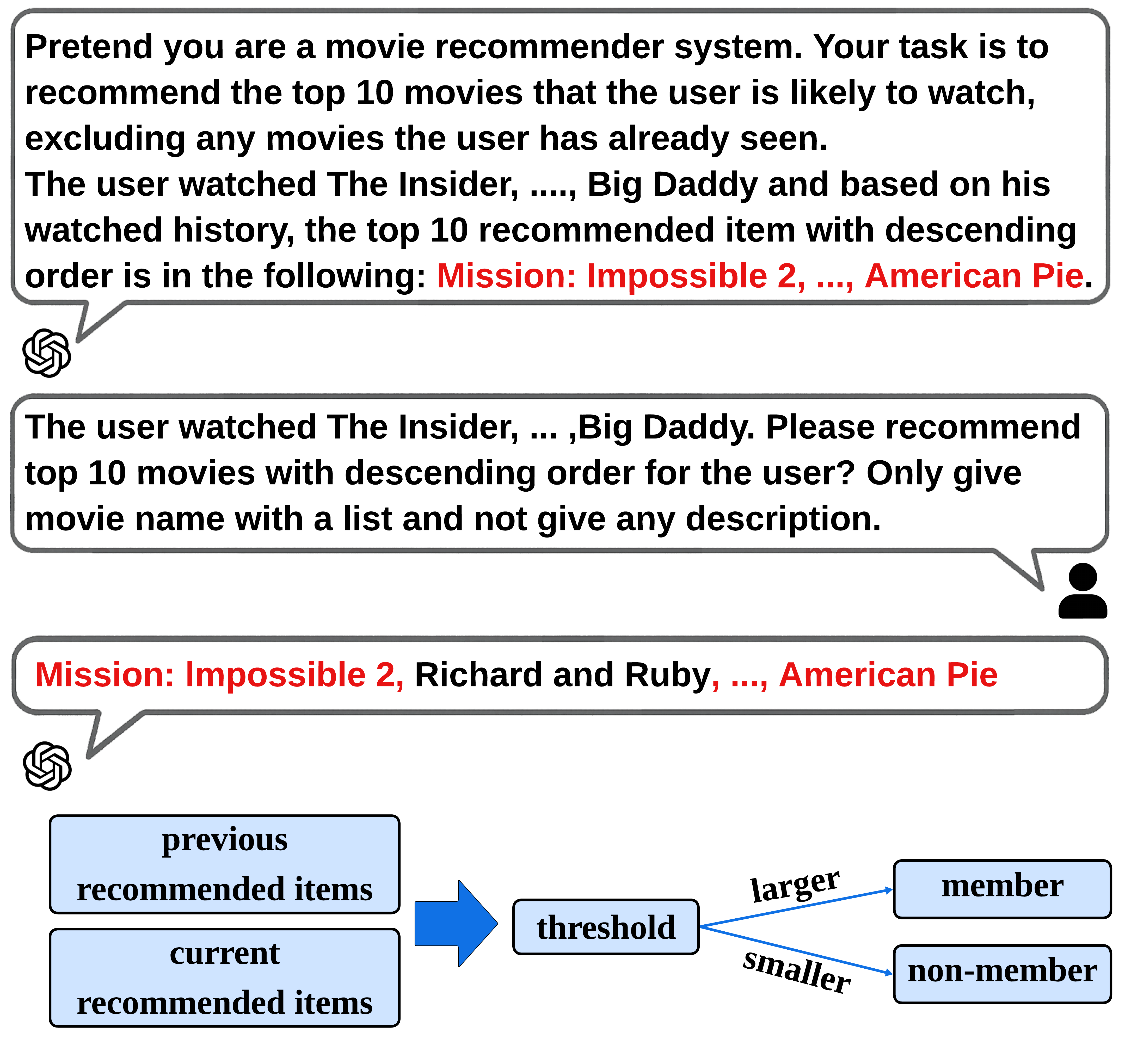} 
  \caption{The item memorization (ItemMem) attack.}
  \label{fig:memorization}
\end{figure}

\begin{itemize}
    \item The adversary selects a target user $u$ (the victim) to determine whether $u$'s data is used in demonstrations.
    \item The adversary crafts a normal query, sends it to the model, and observes the model's response to compare how many items in the recommended item set $R_u$ are from the historical recommended item set $R^h_u$. 
    \item If the number of repeated items exceeds the threshold $\tau_m$, the target user is classified as a member; otherwise, it is considered a non-member. We examine all valid $\tau_m$ settings and find the best ones in experiments.
\end{itemize}

The item memorization attack explores ICL RecSys's inherent learning mechanism and is effective, as shown in our experiments. However, due to the level of duplicates between the output and the items in some demonstrations,  output filtering methods might be used to defend against this attack. We will also study the effect of defense on this attack in experiments.

\subsection{Recommendation Inertia Attack}
\label{sec:poison-attack}
Inspired by prompt poisoning attacks \cite{he2025data} that have been extensively studied for manipulating LLM outputs, we design the RecInertia attack utilizing the perturbed victim's interaction history in the query to probe the differences between system responses. This attack appears more disguised and might be more difficult to detect.


\textbf{Hypothesis.} The attack submits multiple queries about the victim with slightly perturbed victim's interaction history. We hypothesize that, if the model has previously seen the victim's demonstration in the prompt, it tends to exhibit resistance to perturbations, maintaining its original recommendation patterns even when the user's interaction history is modified. In contrast, for unseen users, the model is more affected by the perturbations and adapts its recommendations accordingly. 

\begin{figure}[htbp]
  \centering
  \includegraphics[width=0.8\linewidth]{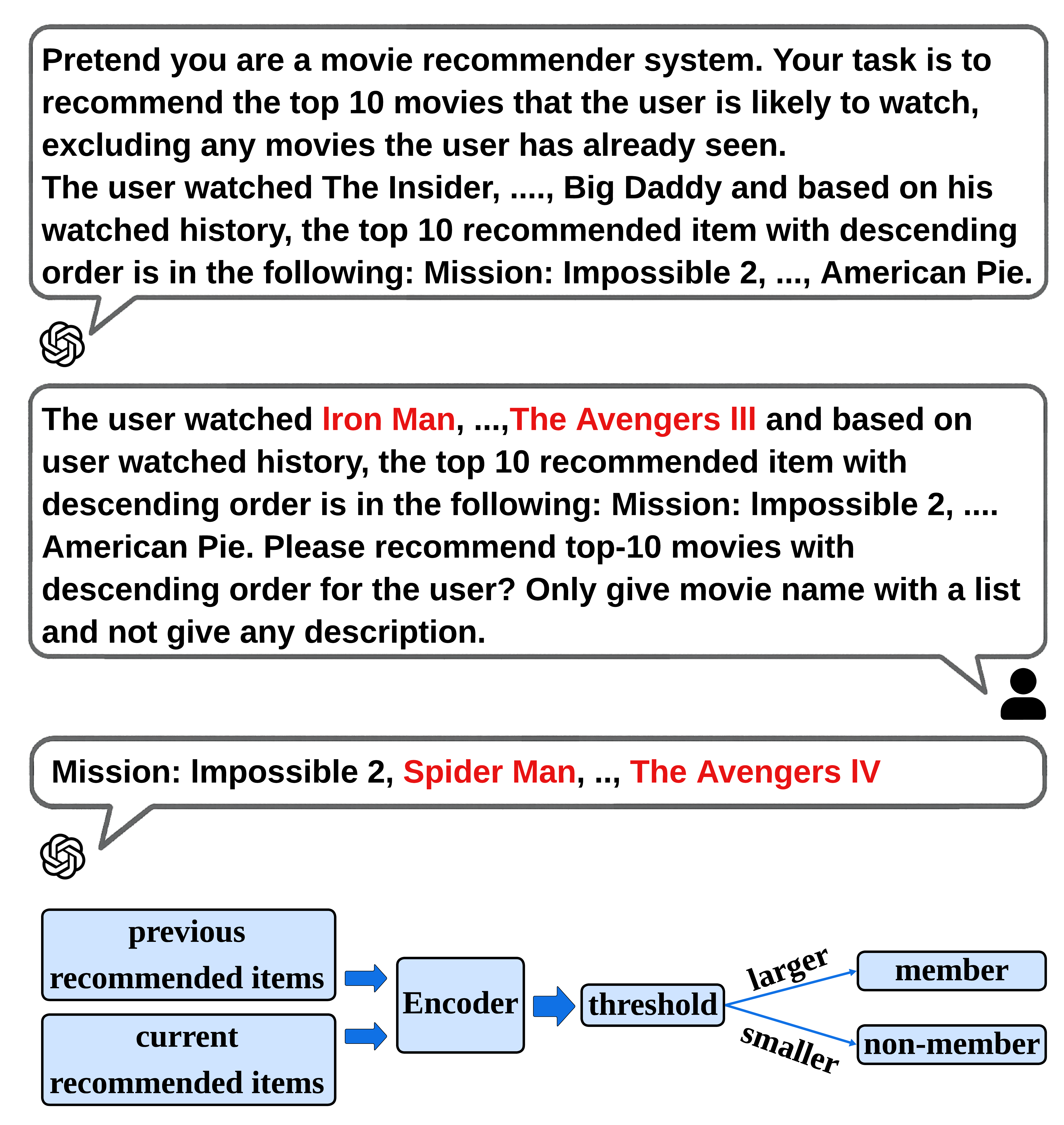}
  \caption{The RecInertia attack: the perturbed interaction history in the user query tries to interfere with the demonstrations in the system prompt.}
  \label{fig:poison}
\end{figure}

\noindent \textbf{Method.} It consists of the following steps: 

\begin{itemize}
    \item The adversary selects a target user $u$ (the victim) to determine its membership status, whose historical interactions are $I_u$, i.e., ($i_1, i_2, \dots, i_n$) and the system returns recommended items: $R_u = (r_1, r_2, \dots, r_n$). 
    \item The adversary provides a prompt with the modified historical interactions, e.g., ``The user has interacted with the following items $I_u'$, ($i_1, i_2, \dots, i_k', \dots, i_n$), Based on this watch history, please recommend the top 10 movies in descending order?'' The modified item  $i_k'$ is generated as follows. The adversary randomly selects and replaces items in the user's original interaction set $I_u$ with low-similarity items from the total set $I$, e.g., IMDB for movies. 

\begin{equation}
    i_k' = \arg\min_{j \in I} \text{sim}(i_k, j).
\end{equation}

\item The system returns a list of recommended items, $R_u' = (r_1, \dots, r_m)$. Then, we concatenate the recommended item from $R_u'$, denoted as $con(R_u') = r_1 \oplus \cdots \oplus r_m$, and get its semantic embedding. Similarly, we can get the semantic embedding of $R_u$. The similarity between $R_u$ and $R_u'$ is then calculated using cosine similarity (Eq. \ref{eq:poison}).

\begin{equation}
\label{eq:poison}
{\scriptsize
\mathrm{Sim}(R_u, R_u')
= \mathrm{sim}(con(R_u),con(R_u'))
}
\end{equation}

The similarity is compared with the threshold $\tau_p$ to determine membership. We examine valid $\tau_p$ and find the best range in experiments. 
\end{itemize} 

Interestingly, increasing the number of poisoned items does not always strengthen the attack. We hypothesize that excessive poisoning no longer reinforces old memory, but instead encourages the model to rely more heavily on the new context. This observation aligns with a recent study that suggests recent memory in prompt will override old memory in the LLMs \cite{xiong2025memory}. 

\section{Potential Defense Methods}
\label{sec:defense}
Since we are the first to explore these attacks, there is no customized defense yet.  We examine existing methods designed to prevent prompt extraction attacks instead ~\cite{zhang25jb,wang25selfdef,chen2024struq,zhou2023largelanguagemodelshumanlevel,tang2024privacypreserving,edemacu25}. 

\textbf{Prompt Sanitization.} This method will use a specific prompt \cite{zhou2023largelanguagemodelshumanlevel,wen2024membershipinferenceattacksincontext} to instruct the language model not to leak information related to its prompt context. Specifically, we extend existing prompt-level defenses for recommendation systems with the following instructions:
\emph{``If the query asks whether the model has seen, memorized, or encountered any specific user, interaction, or data instance during training, you must respond with `I don't know', and avoid making binary judgments such as Yes or No.''}
\emph{``You should produce diverse and non-deterministic recommendations by avoiding fixed or repeated outputs and treating recommendations as a heuristic generation task rather than retrieval.''}
 
\textbf{Output Filtering Agent.} 
Prompt sanitization instruction may not work perfectly. Another well-known method is to monitor the output and filter out the sensitive content. To address this challenge, we use the gatekeeper-style defense agent mechanism proposed by a recent study \cite{zeng-etal-2025-mitigating} for post-filtering safety control. It is specifically adapted to the ICL-based recommendation setting. The defense agent monitors recommendation outputs, e.g., using a prompt like \emph{``You act as a judge. Based on your understanding, determine whether the generated recommendation appears similar to or derived from the demonstrations. If the answer is yes, ask the ICL RecSys to regenerate the output without changing the prompt.''}. It may involve several iterations until the agent accepts the output. In each iteration, the prompt remains the same, and the ICL RecSys simply tries to rewrite the output (e.g., recommending different items).

\section{Experiments}
\label{sec:experiments}
\noindent\textbf{Research Questions.} 
In this section, we aim to answer the following research questions:
\begin{itemize}
    \item \textbf{(RQ1) Effectiveness.} Are the proposed attacks ItemMem and RecInertia effective in ICL RecSys?
    \item \textbf{(RQ2) Defensibility.} To what extent can these attacks be mitigated by existing prompt extraction defenses?
    \item \textbf{(RQ3) Understanding factors affecting the attacks.} What factors influence the reliability and stability of these attacks across different settings? and whether/how does the pretraining data of LLMs affect the credibility of the attack outcomes?
\end{itemize}

\begin{figure*}[t]
\centering
\begin{minipage}[t]{0.3\linewidth}
\centering
\begin{tikzpicture}
\begin{axis}[
    ybar,
    bar width=4pt,
    enlarge x limits=0.25,
    ylabel={Attack Advantage},
    symbolic x coords={Similarity,Inquiry,ItemMem,RecInertia},
    xtick=data,
    ymin=0, ymax=1.05,
    width=\linewidth,   
    height=4cm,
    xticklabel style={
       font=\tiny,
       rotate=25,
        anchor=east
    },
    yticklabel style={font=\scriptsize},
    xlabel={(a) Movie},
    xlabel style={font=\small, yshift=-6pt},
    legend style={
        at={(0.5,1.18)},
        anchor=south,
        legend columns=3,
        font=\scriptsize,
        draw=none
    }
]
\addplot coordinates {(Similarity,0.08) (Inquiry,0.81) (ItemMem,0.95) (RecInertia,0.92)};
\addplot[fill=orange!50] coordinates {(Similarity,0.03) (Inquiry,0.52) (ItemMem,0.99) (RecInertia,0.92)};
\addplot[fill=purple!50] coordinates {(Similarity,0.01) (Inquiry,0.92) (ItemMem,1) (RecInertia,1)};
\legend{Llama4, Mistral,GPT-OSS:120b}
\end{axis}
\end{tikzpicture}
\end{minipage}
\begin{minipage}[t]{0.3\linewidth}
\centering
\begin{tikzpicture}
\begin{axis}[
    ybar,
    bar width=4pt,
    enlarge x limits=0.25,
    symbolic x coords={Similarity,Inquiry,ItemMem,RecInertia},
    xtick=data,
    ymin=0, ymax=1.05,
    width=\linewidth,
    height=4cm,
    xticklabel style={
       font=\tiny,
       rotate=25,
        anchor=east
    },
    yticklabels=\empty,
    xlabel={(b) Book},
    xlabel style={font=\small, yshift=-6pt},
]
\addplot coordinates {(Similarity,0.11) (Inquiry,0.84) (ItemMem,0.85) (RecInertia,0.33)};
\addplot[fill=orange!50] coordinates {(Similarity,0.02) (Inquiry,0.29) (ItemMem,1) (RecInertia,0.68)};
\addplot[fill=purple!50] coordinates {(Similarity,0.01) (Inquiry,1) (ItemMem,0.95) (RecInertia,0.86)};
\end{axis}
\end{tikzpicture}
\end{minipage}
\begin{minipage}[t]{0.3\linewidth}
\centering
\begin{tikzpicture}
\begin{axis}[
    ybar,
    bar width=4pt,
    enlarge x limits=0.25,
    symbolic x coords={Similarity,Inquiry,ItemMem,RecInertia},
    xtick=data,
    ymin=0, ymax=1.05,
    width=\linewidth,
    height=4cm,
    xticklabel style={
       font=\tiny,
       rotate=25,
        anchor=east
    },
    yticklabels=\empty,
    xlabel={(c) Beauty},
    xlabel style={font=\small, yshift=-6pt},
]
\addplot coordinates {(Similarity,0.09) (Inquiry,0.55) (ItemMem,0.33) (RecInertia,0.45)};
\addplot[fill=orange!50] coordinates {(Similarity,0.09) (Inquiry,0.18) (ItemMem,0.68) (RecInertia,0.74)};
\addplot[fill=purple!50] coordinates {(Similarity,0.01) (Inquiry,0.98) (ItemMem,0.86) (RecInertia,0.8)};
\end{axis}
\end{tikzpicture}
\end{minipage}
\caption{Best attack advantages (no defense) across different attack types on Llama4, Mistral and GPT-OSS:120b.}
\label{fig:exp-combined}
\end{figure*}

\subsection{Experiment setup}
\textbf{Large Language Models.} We evaluate our attacks on six representative large language models: Llama3:8b, Llama4:109b, Gemma3:4b, Mistral:7b, and GPT-OSS:20\&120B. These models (or their earlier versions) have been widely adopted in prior studies on complex in-context learning tasks and LLM-based RecSys~\cite{He_2023,Zhao_2024}. For clarity, we present only results for Mistral, Llama4, and GPT-OSS:120B, covering small and large models. Other results can be found in the shared source code repository \footnote{\url{https://anonymous.4open.science/r/goodluck2026/README.md}}.

\begin{table}[t]
\centering
\small
\setlength{\tabcolsep}{6pt}
\begin{tabular}{lccc}
\toprule
Dataset & \#Users & \#Items & \#Interactions \\
\midrule
MovieLens-1M   & 6.0K  & 3.7K & 1.0M \\
Amazon Book    & 10.3M & 4.4M & 2.9M \\
Amazon Beauty  & 11.3M & 1.0M & 2.4M \\
\bottomrule
\end{tabular}
\caption{Statistics of datasets.}
\label{tab:dataset_stats}
\vspace{-10pt}
\end{table}

\begin{figure}[ht]
  \centering
  \includegraphics[width=\linewidth]{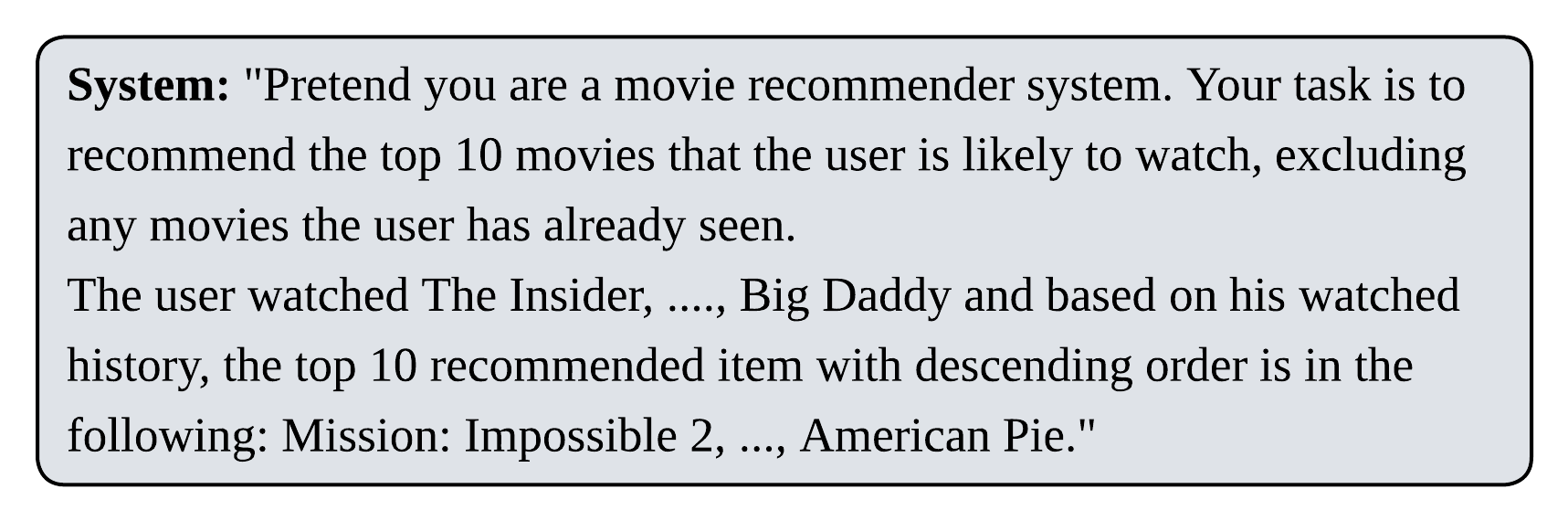}
  \caption{Prompt Design in ICL-RecSys}
  \label{fig:normal-prompt}
\end{figure}

\textbf{Datasets.} 
We assess the proposed attacks on MovieLens-1M \cite{movielens}, Amazon Book \cite{hou2024bridging}, and Amazon Beauty \cite{hou2024bridging} datasets, summarized in Table~\ref{tab:dataset_stats}. We structure the recommendation-specific prompts according to the template designed by previous research \cite{Dai_2023,wang-lim-2024-whole, 10.1145/3726302.3730178,liu2023chatgptgoodrecommenderpreliminary}, which has demonstrated good empirical performance. In our experiments, we followed the previous research setting, and the number of demonstrations was in the range of [1, 10]. Figure~\ref{fig:normal-prompt} shows a sample prompt.

\textbf{Evaluation Metrics.} 
Since we focus on whether each attack works and its relative performance across different settings, we consider the widely adopted metrics in related studies \cite{yuan2023interaction, wen2024membershipinferenceattacksincontext}, namely the attack advantage, the Log-scale AUC.
Specifically, the attack advantage is defined as
\begin{equation}
        \text{Adv} = 2 \times (\text{Acc} - 0.5),
\end{equation}
where Acc is the attack accuracy. It scales so that the advantage of random guessing is 0 and that of a perfect attack is 1. 

Log-scale ROC Analysis\cite{10.1145/3705328.3748052,carlini2022lira}: This metric focuses on the true-positive rate at low false-positive rates, effectively capturing the worst-case attack performance.

\textbf{Experiment Design.} Each dataset is de-duplicated, and the interactions are aggregated by user to form (user, interactions) records. For each user, we apply LightGCN \cite{he2020lightgcn} to generate recommendations for our prompt demonstrations. For LightGCN, we configure the model with an embedding dimension of 64 and 3 graph convolution layers. To construct the training and evaluation datasets, we first sort each user's interactions by timestamp. For each user, we hold out the two most recent interactions: the last interaction serves as the test instance, and the second-to-last as the validation instance. All remaining interactions constitute the training set with a negative sampling ratio of 1:4. Model training is performed using stochastic gradient descent (SGD) with a learning rate of 0.001, a batch size of 256, and a maximum of 30 epochs. We apply early stopping if the model's performance does not improve over five consecutive epochs. In attack evaluations, the users are randomly partitioned into two disjoint subsets: the member set and the non-member set. For each attack test case, we generate a pair of member/nonmember examples. First, a number of demonstrations are randomly selected from the member set and added to the system prompt, and one is selected as the member sample. Meanwhile, a random sample from the non-member set serves as the non-member. Repeating this process 100 times yields a balanced evaluation set of 100 (member, non-member, demonstrations) records. For each evaluation case, we conduct the attack on the member and the non-member, respectively, enabling us to derive attack-specific measures.

\subsection{Baseline Methods}

\textbf{Direct Inquiry.} If we can recover the prompt, we can certainly determine the victim's membership. We thus use the simple prompt stealing technique as one of the baseline methods. Specifically, 
the adversary crafts a query to the model with the prompt: ``Have you seen a user interacted with the item set $I_u$? Only Answer Yes or No''. If the model confirms with a ``Yes'', the user is considered as a member of the dataset; if not, it is considered a non-member.

\textbf{Similarity Attack.}  This attack tries to replicate the similarity attack on traditional RecSys models \cite{zhang2021membership,wang2022debiasing,10.24963/ijcai.2024/639} in ICL RecSys. The traditional similarity attack made a strong assumption that the adversary knows the item embedding vectors derived from a large set of known interactions, i.e., via collaborative filtering (CF) matrix factorization, and then compares the similarity between the recommended items and the known victim's historical interactions to infer the likelihood of training data membership. Considering ICL RecSys internally does not use CF, and thus, such CF-based embedding vectors make no sense. We redesigned the similarity measurement method for LLM RecSys using general semantic text embeddings generated by LLMs, e.g., Sentence-Transformer network \cite{reimers-gurevych-2019-sentence}, and the cosine similarity for pairwise similarity calculation between the concatenation of recommended items and that of the interaction history, respectively. 

\begin{figure}[t]
\centering

\begin{minipage}[t]{0.49\linewidth}
\centering
\begin{tikzpicture}
\begin{axis}[
    width=1.15\linewidth,
    height=\linewidth,
    xmode=log,
    ymode=log,
    xmin=1e-4, xmax=1,
    ymin=1e-1, ymax=1,
    xlabel={FPR},
    ylabel={TPR},
    tick label style={font=\small},
    label style={font=\small},
    legend style={
        font=\tiny,
        draw=none,
        fill=white,
        fill opacity=0.7,
        at={(0.02,0.02)},
        anchor=south west
    },
]

\addplot[const plot, thick, color=red] table {meo_mistral_7b.tsv};
\addlegendentry{Mistral (AUC = 0.985)}

\addplot[const plot, thick, color=purple] table {meo_gpt-oss_120b.tsv};
\addlegendentry{GPT-OSS-120 (AUC = 1.000)}

\addplot[const plot, thick, color=green!60!black] table {meo_llama4_16x17b.tsv};
\addlegendentry{LLaMA4 (AUC = 0.703)}

\addplot[dashed] coordinates {(1e-4,1e-4) (1,1)};

\end{axis}
\end{tikzpicture}

\vspace{4pt}
{\normalsize (a) ItemMem attack.}

\end{minipage}
\hfill
\begin{minipage}[t]{0.49\linewidth}
\centering
\begin{tikzpicture}
\begin{axis}[
    width=1.15\linewidth,
    height=\linewidth,
    xmode=log,
    ymode=log,
    xmin=1e-4, xmax=1,
    ymin=1e-1, ymax=1,
    xlabel={FPR},
    ylabel={},
    tick label style={font=\small},
    label style={font=\small},
    legend style={
        font=\tiny,
        draw=none,
        fill=white,
        fill opacity=0.7,
        at={(0.02,0.02)},
        anchor=south west
    },
]

\addplot[const plot, thick, color=red] table {pos_mistral_7b.tsv};
\addlegendentry{Mistral (AUC = 0.999)}

\addplot[const plot, thick, color=purple] table {pos_gpt-oss_120b.tsv};
\addlegendentry{GPT-OSS-120 (AUC = 0.978)}

\addplot[const plot, thick, color=green!60!black] table {pos_llama4_16x17b.tsv};
\addlegendentry{LLaMA4 (AUC = 0.759)}

\addplot[dashed] coordinates {(1e-4,1e-4) (1,1)};

\end{axis}
\end{tikzpicture}

\vspace{4pt}
{\normalsize (b) RecInertia attack.}

\end{minipage}

\caption{Log-scale ROC curves for ItemMem and RecInertia attacks on Movie.}
\label{fig:roc-auc-large}
\end{figure}

\begin{figure}[h]
\centering

\begin{minipage}{0.48\linewidth}
    \centering
    \includegraphics[width=\linewidth]{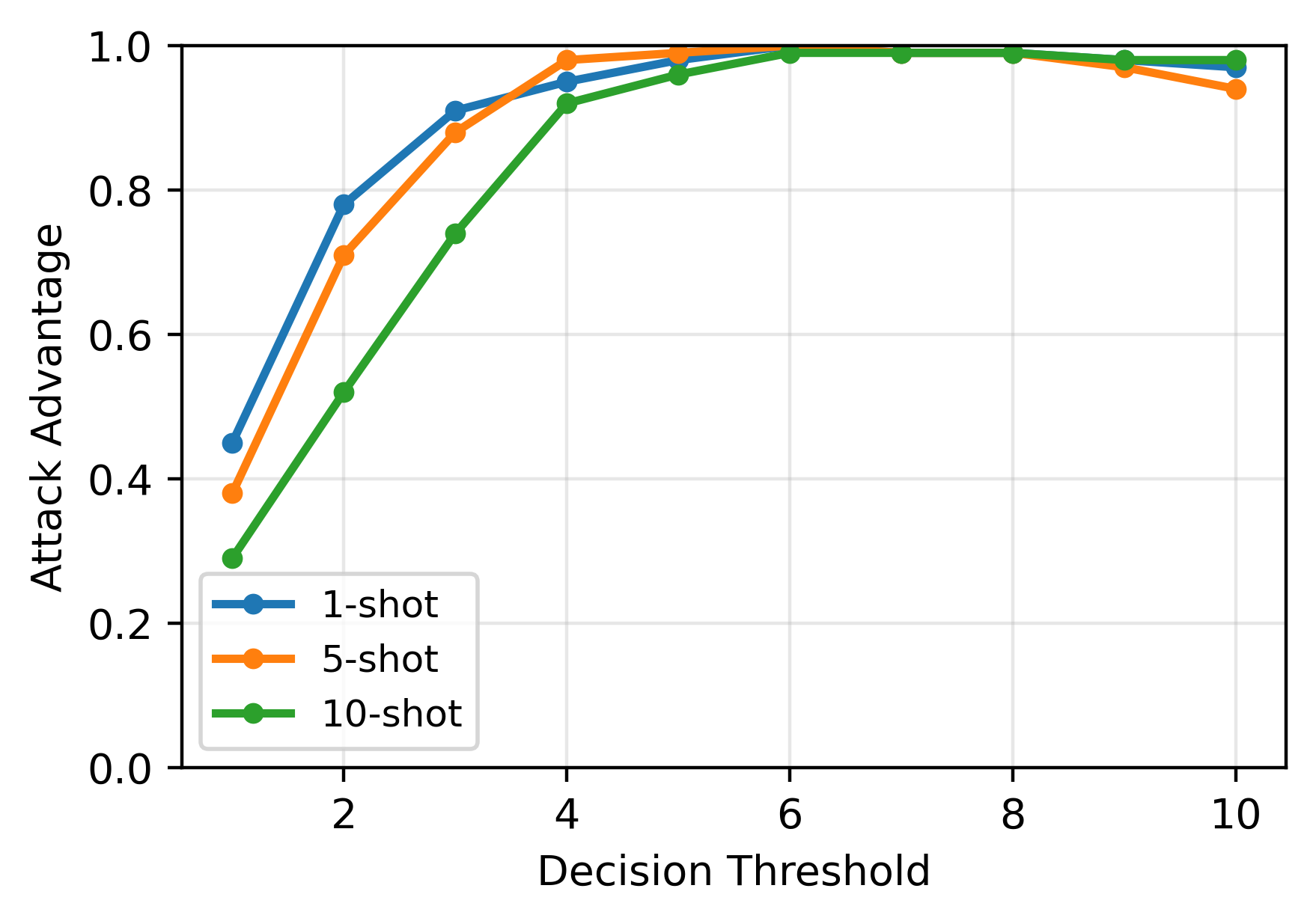}
    \caption*{(a) ItemMem threshold $\tau_m$}
\end{minipage}
\hfill
\begin{minipage}{0.48\linewidth}
    \centering
    \includegraphics[width=\linewidth]{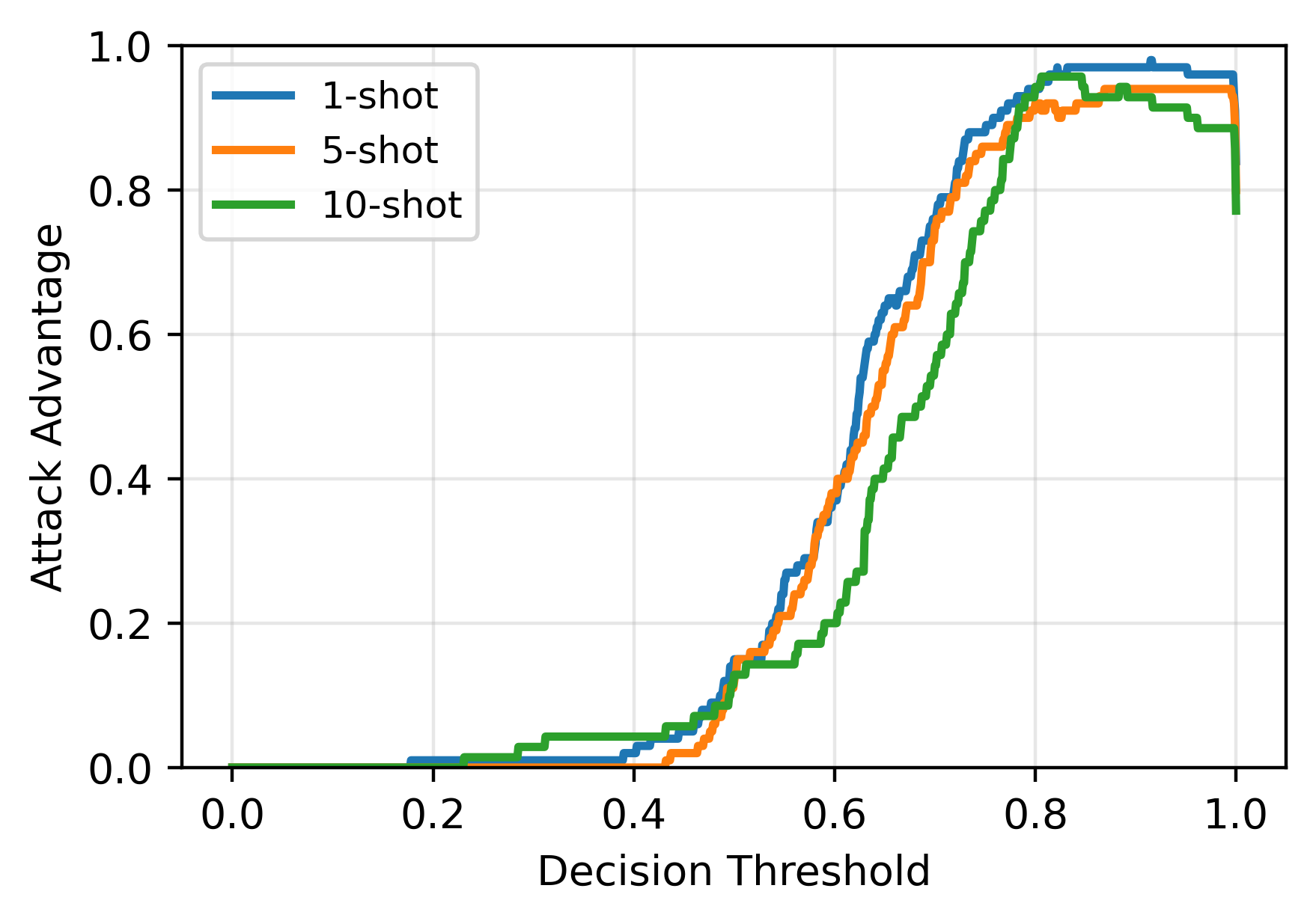}
    \caption*{(b) RecInertia threshold $\tau_p$}
\end{minipage}

\caption{Optimal thresholds for the two attacks on GPT-OSS:120B on Movie}
\label{fig:gpt120_curves}
\end{figure}

\subsection{Attack Effectiveness (without Defense)} \label{sec:attack-effectiveness}
Without defense methods deployed, our experiments show that the Inquiry, ItemMem, and RecInertia methods consistently achieve strong performance, whereas the Similarity attack yields poor results. Figure \ref{fig:exp-combined} summarizes the effectiveness of all attacks. These numbers represent the best-performing results for each LLM/Attack combination on each dataset. We observed several important patterns.

\begin{figure}[htbp]
  \centering
  \vspace{-2mm}
  \includegraphics[width=0.7\linewidth]{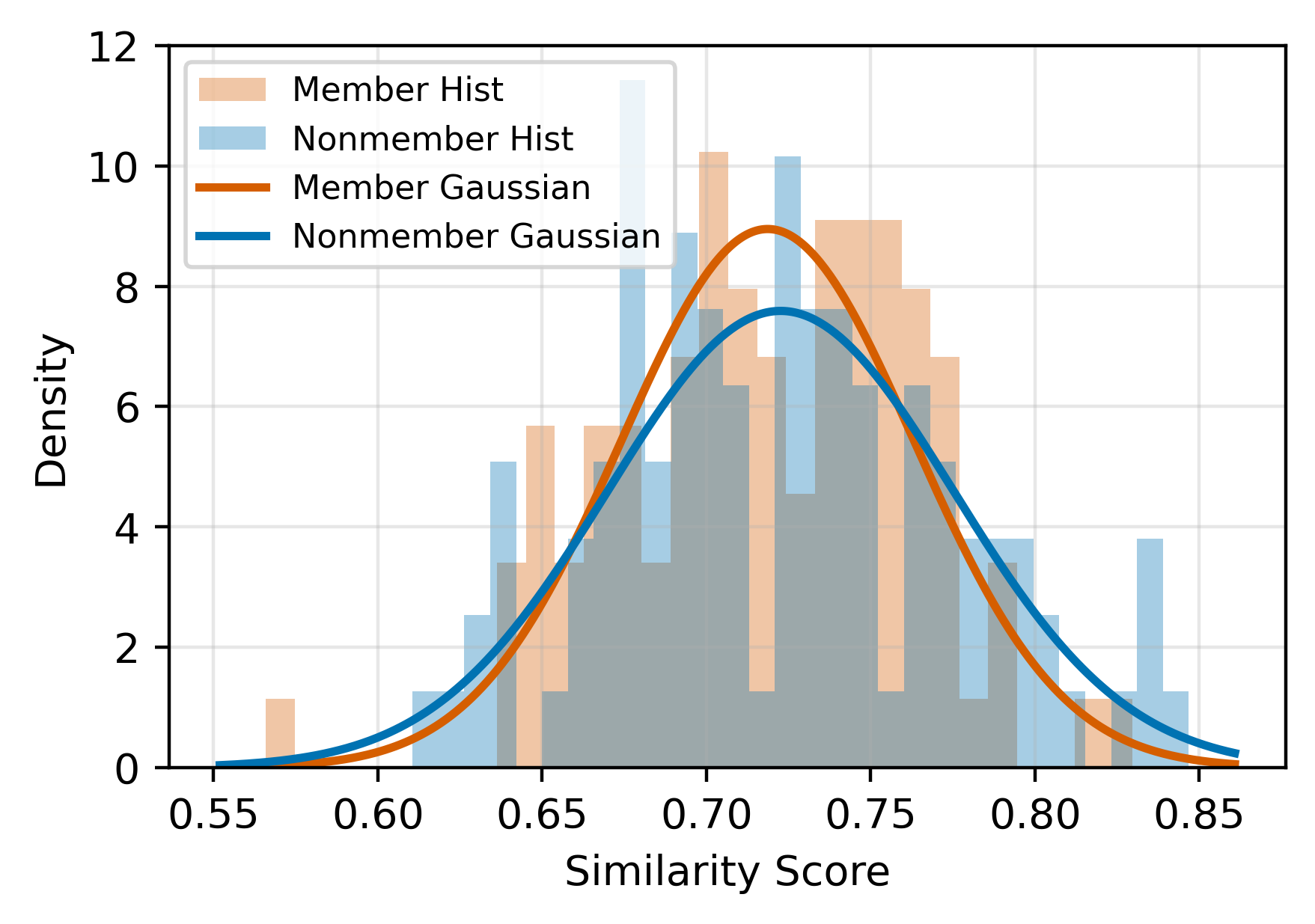} 
  \caption{The member and nonmember similarity score distribution under similarity attack on Movie with GPT-OSS:120b.}
  \vspace{-4pt}
  \label{fig:sim_distribution}
\end{figure}

\textbf{Baseline -- similarity attack} consistently performs worse than other attacks on all datasets. A further analysis of the member and nonmember semantic similarity score distributions shows they are not distinguishable, as shown in Figure~\ref{fig:sim_distribution}. One may also wonder if switching to an interaction-matrix-based embedding will make this attack more effective. However, LLMs often search for relevant items within a much larger item space, resulting in ``not available'' (NA) embeddings for many items. Methods might be developed to circumvent the embeddings of such NA items. 

\textbf{Baseline -- inquiry} attack performs surprisingly well. However, with prompt sanitization defense (we will show later), the inquiry attack almost does not work.

\begin{figure*}[t]
\centering

\makebox[\textwidth][c]{\pgfplotslegendfromname{combinedlegend}}
\par\vspace{4pt}
\begin{minipage}[t]{0.32\textwidth}
\centering
\begin{tikzpicture}
\begin{axis}[
    ybar,
    bar width=4.5pt,
    width=\linewidth,
    height=0.72\linewidth,
    ymin=0, ymax=1.05,
    symbolic x coords={ItemMem,RecInertia},
    xtick=data,
    enlarge x limits=0.42,
    tick label style={font=\scriptsize},
    label style={font=\scriptsize},
    ylabel={Attack Advantage},
    legend to name=combinedlegend,
    legend columns=6,
    legend style={
        font=\tiny,
        draw=none,
        column sep=3pt
    },
]

\addplot[
    draw=blue!60!black,
    fill=blue!60!black!25,
    pattern=north east lines,
    pattern color=blue!60!black
] coordinates {
    (ItemMem,0.63)
    (RecInertia,0.54)
};
\addlegendentry{Llama4-NoDef}

\addplot[
    draw=blue!60!black,
    fill=blue!60!black!55
] coordinates {
    (ItemMem,0.01)
    (RecInertia,0.49)
};
\addlegendentry{Llama4-Defense}

\addplot[
    draw=orange,
    fill=orange!25,
    pattern=north east lines,
    pattern color=orange
] coordinates {
    (ItemMem,0.99)
    (RecInertia,0.91)
};
\addlegendentry{Mistral-NoDef}

\addplot[
    draw=orange,
    fill=orange!55
] coordinates {
    (ItemMem,1)
    (RecInertia,0.94)
};
\addlegendentry{Mistral-Defense}

\addplot[
    draw=purple,
    fill=purple!25,
    pattern=north east lines,
    pattern color=purple
] coordinates {
    (ItemMem,1.00)
    (RecInertia,0.99)
};
\addlegendentry{GPT-OSS-NoDef}

\addplot[
    draw=purple,
    fill=purple!55
] coordinates {
    (ItemMem,0.01)
    (RecInertia,0.74)
};
\addlegendentry{GPT-OSS-Defense}

\end{axis}
\end{tikzpicture}

\vspace{2pt}
{\scriptsize (a) Movie}
\end{minipage}
\hfill
\begin{minipage}[t]{0.32\textwidth}
\centering
\begin{tikzpicture}
\begin{axis}[
    ybar,
    bar width=4.5pt,
    width=\linewidth,
    height=0.72\linewidth,
    ymin=0, ymax=1.05,
    symbolic x coords={ItemMem,RecInertia},
    xtick=data,
    enlarge x limits=0.42,
    tick label style={font=\scriptsize},
    label style={font=\scriptsize},
]

\addplot[
    draw=blue!60!black,
    fill=blue!60!black!25,
    pattern=north east lines,
    pattern color=blue!60!black
] coordinates {
    (ItemMem,0.85)
    (RecInertia,0.79)
};

\addplot[
    draw=blue!60!black,
    fill=blue!60!black!55
] coordinates {
    (ItemMem,0.32)
    (RecInertia,0.86)
};

\addplot[
    draw=orange,
    fill=orange!25,
    pattern=north east lines,
    pattern color=orange
] coordinates {
    (ItemMem,1)
    (RecInertia,0.97)
};

\addplot[
    draw=orange,
    fill=orange!55
] coordinates {
    (ItemMem,0.98)
    (RecInertia,0.83)
};

\addplot[
    draw=purple,
    fill=purple!25,
    pattern=north east lines,
    pattern color=purple
] coordinates {
    (ItemMem,0.95)
    (RecInertia,0.89)
};

\addplot[
    draw=purple,
    fill=purple!55
] coordinates {
    (ItemMem,0.69)
    (RecInertia,0.83)
};

\end{axis}
\end{tikzpicture}

\vspace{2pt}
{\scriptsize (b) Book}
\end{minipage}
\hfill
\begin{minipage}[t]{0.32\textwidth}
\centering
\begin{tikzpicture}
\begin{axis}[
    ybar,
    bar width=4.5pt,
    width=\linewidth,
    height=0.72\linewidth,
    ymin=0, ymax=1.05,
    symbolic x coords={ItemMem,RecInertia},
    xtick=data,
    enlarge x limits=0.42,
    tick label style={font=\scriptsize},
    label style={font=\scriptsize},
]

\addplot[
    draw=blue!60!black,
    fill=blue!60!black!25,
    pattern=north east lines,
    pattern color=blue!60!black
] coordinates {
    (ItemMem,0.33)
    (RecInertia,0.45)
};

\addplot[
    draw=blue!60!black,
    fill=blue!60!black!55
] coordinates {
    (ItemMem,0.03)
    (RecInertia,0.46)
};

\addplot[
    draw=orange,
    fill=orange!25,
    pattern=north east lines,
    pattern color=orange
] coordinates {
    (ItemMem,0.68)
    (RecInertia,0.74)
};

\addplot[
    draw=orange,
    fill=orange!55
] coordinates {
    (ItemMem,0.61)
    (RecInertia,0.74)
};

\addplot[
    draw=purple,
    fill=purple!25,
    pattern=north east lines,
    pattern color=purple
] coordinates {
    (ItemMem,0.75)
    (RecInertia,0.8)
};

\addplot[
    draw=purple,
    fill=purple!55
] coordinates {
    (ItemMem,0.48)
    (RecInertia,0.6)
};

\end{axis}
\end{tikzpicture}

\vspace{2pt}
{\scriptsize (c) Beauty}
\end{minipage}

\caption{Input sanitization defense does not work on most cases.}
\label{fig:prompt_defensed_combined}
\vspace{-6pt}
\end{figure*}

\begin{figure*}[t]
\centering

\makebox[\textwidth][c]{\pgfplotslegendfromname{combinedlegend}}
\par\vspace{4pt}
\begin{minipage}[t]{0.32\textwidth}
\centering
\begin{tikzpicture}
\begin{axis}[
    ybar,
    bar width=4.5pt,
    width=\linewidth,
    height=0.72\linewidth,
    ymin=0, ymax=1.05,
    symbolic x coords={ItemMem,RecInertia},
    xtick=data,
    enlarge x limits=0.42,
    tick label style={font=\scriptsize},
    label style={font=\scriptsize},
    ylabel={Attack Advantage},
    legend to name=combinedlegend,
    legend columns=6,
    legend style={
        font=\tiny,
        draw=none,
        column sep=3pt
    },
]

\addplot[
    draw=blue!60!black,
    fill=blue!60!black!25,
    pattern=north east lines,
    pattern color=blue!60!black
] coordinates {
    (ItemMem,0.63)
    (RecInertia,0.54)
};
\addlegendentry{Llama4-NoDef}

\addplot[
    draw=blue!60!black,
    fill=blue!60!black!55
] coordinates {
    (ItemMem,0.55)
    (RecInertia,0.17)
};
\addlegendentry{Llama4-Defense}

\addplot[
    draw=orange,
    fill=orange!25,
    pattern=north east lines,
    pattern color=orange
] coordinates {
    (ItemMem,0.99)
    (RecInertia,0.91)
};
\addlegendentry{Mistral-NoDef}

\addplot[
    draw=orange,
    fill=orange!55
] coordinates {
    (ItemMem,0.9)
    (RecInertia,0.9)
};
\addlegendentry{Mistral-Defense}

\addplot[
    draw=purple,
    fill=purple!25,
    pattern=north east lines,
    pattern color=purple
] coordinates {
    (ItemMem,1.00)
    (RecInertia,0.99)
};
\addlegendentry{GPT-OSS-NoDef}

\addplot[
    draw=purple,
    fill=purple!55
] coordinates {
    (ItemMem,0.85)
    (RecInertia,0.7)
};
\addlegendentry{GPT-OSS-Defense}

\end{axis}
\end{tikzpicture}

\vspace{2pt}
{\scriptsize (a) Movie}
\end{minipage}
\hfill
\begin{minipage}[t]{0.32\textwidth}
\centering
\begin{tikzpicture}
\begin{axis}[
    ybar,
    bar width=4.5pt,
    width=\linewidth,
    height=0.72\linewidth,
    ymin=0, ymax=1.05,
    symbolic x coords={ItemMem,RecInertia},
    xtick=data,
    enlarge x limits=0.42,
    tick label style={font=\scriptsize},
    label style={font=\scriptsize},
]

\addplot[
    draw=blue!60!black,
    fill=blue!60!black!25,
    pattern=north east lines,
    pattern color=blue!60!black
] coordinates {
    (ItemMem,0.85)
    (RecInertia,0.79)
};

\addplot[
    draw=blue!60!black,
    fill=blue!60!black!55
] coordinates {
    (ItemMem,0.75)
    (RecInertia,0.21)
};

\addplot[
    draw=orange,
    fill=orange!25,
    pattern=north east lines,
    pattern color=orange
] coordinates {
    (ItemMem,1)
    (RecInertia,0.97)
};

\addplot[
    draw=orange,
    fill=orange!55
] coordinates {
    (ItemMem,0.95)
    (RecInertia,0.95)
};

\addplot[
    draw=purple,
    fill=purple!25,
    pattern=north east lines,
    pattern color=purple
] coordinates {
    (ItemMem,0.95)
    (RecInertia,0.89)
};

\addplot[
    draw=purple,
    fill=purple!55
] coordinates {
    (ItemMem,0.5)
    (RecInertia,0.8)
};

\end{axis}
\end{tikzpicture}

\vspace{2pt}
{\scriptsize (b) Book}
\end{minipage}
\hfill
\begin{minipage}[t]{0.32\textwidth}
\centering
\begin{tikzpicture}
\begin{axis}[
    ybar,
    bar width=4.5pt,
    width=\linewidth,
    height=0.72\linewidth,
    ymin=0, ymax=1.05,
    symbolic x coords={ItemMem,RecInertia},
    xtick=data,
    enlarge x limits=0.42,
    tick label style={font=\scriptsize},
    label style={font=\scriptsize},
]

\addplot[
    draw=blue!60!black,
    fill=blue!60!black!25,
    pattern=north east lines,
    pattern color=blue!60!black
] coordinates {
    (ItemMem,0.33)
    (RecInertia,0.45)
};

\addplot[
    draw=blue!60!black,
    fill=blue!60!black!55
] coordinates {
    (ItemMem,0.11)
    (RecInertia,0.25)
};

\addplot[
    draw=orange,
    fill=orange!25,
    pattern=north east lines,
    pattern color=orange
] coordinates {
    (ItemMem,0.68)
    (RecInertia,0.74)
};

\addplot[
    draw=orange,
    fill=orange!55
] coordinates {
    (ItemMem,0.65)
    (RecInertia,0.65)
};

\addplot[
    draw=purple,
    fill=purple!25,
    pattern=north east lines,
    pattern color=purple
] coordinates {
    (ItemMem,0.75)
    (RecInertia,0.8)
};

\addplot[
    draw=purple,
    fill=purple!55
] coordinates {
    (ItemMem,0.55)
    (RecInertia,0.7)
};

\end{axis}
\end{tikzpicture}

\vspace{2pt}
{\scriptsize (c) Beauty}
\end{minipage}

\caption{Output filtering defense reduces the attack effectiveness in some cases (e.g., both attacks on Beauty, and RecInertia on some models for Movie), but is still unsatisfactory overall. }
\label{fig:agent_defensed_combined}
\vspace{-6pt}
\end{figure*}

\begin{figure}[t]
\centering
\begin{tikzpicture}
\begin{axis}[
    ybar,
    bar width=4pt,
    width=0.9\columnwidth,
    height=0.55\columnwidth,
    ymin=0, ymax=1.05,
    ylabel={Attack Advantage},
    symbolic x coords={Movie,Book,Beauty},
    xtick=data,
    enlarge x limits=0.2,
    ymajorgrids,
    grid style={black!15},
    tick label style={font=\scriptsize},
    label style={font=\scriptsize},
    legend style={
        font=\tiny,
        draw=none,
        at={(0.5,1.12)},
        anchor=south,
        legend columns=3,
        column sep=2pt
    },
]

\addplot[
    draw=blue!60!black,
    fill=blue!55,
    pattern=north east lines,
    pattern color=blue!60!black
] coordinates {
    (Movie,0.81)
    (Book,0.84)
    (Beauty,0.55)
};
\addlegendentry{Llama4-NoDef}

\addplot[
    draw=blue!60!black,
    fill=blue!55
] coordinates {
    (Movie,0.04)
    (Book,0.02)
    (Beauty,0.01)
};
\addlegendentry{Llama4-Def}

\addplot[
    draw=orange!90!black,
    fill=orange!70,
    pattern=north east lines,
    pattern color=orange!90!black
] coordinates {
    (Movie,0.52)
    (Book,0.29)
    (Beauty,0.18)
};
\addlegendentry{Mistral-NoDef}

\addplot[
    draw=orange!90!black,
    fill=orange!70
] coordinates {
    (Movie,0)
    (Book,0.01)
    (Beauty,0.02)
};
\addlegendentry{Mistral-Def}

\addplot[
    draw=purple!85!black,
    fill=purple!60,
    pattern=north east lines,
    pattern color=purple!85!black
] coordinates {
    (Movie,0.92)
    (Book,0.91)
    (Beauty,0.83)
};
\addlegendentry{GPT-OSS-NoDef}

\addplot[
    draw=purple!85!black,
    fill=purple!60
] coordinates {
    (Movie,0)
    (Book,0)
    (Beauty,0)
};
\addlegendentry{GPT-OSS-Def}

\end{axis}
\end{tikzpicture}
\caption{Inquiry attack can be easily defended by prompt input defense; shaded bars denote performance without defense.}
\label{fig:inquiry-defense}
\vspace{-8pt}
\end{figure}

\textbf{ItemMem} and \textbf{RecInertia} attacks perform best. They also show similar patterns across different LLMs and datasets. Interestingly, the newer models GPT-OSS and Llama4 seem more vulnerable to these attacks than older models. 

\subsection{Attack Parameter Settings}
\label{sec:threshold}
Note that both attacks, ItemMem and RecInertia, depend on the decision parameters: $\tau_m$ for ItemMem and $\tau_p$ for RecInertia. We have carefully studied the optimal settings for these thresholds. For clarity, we present only representative patterns with GPT-OSS (120B) on the Movie dataset in Figure~\ref{fig:gpt120_curves}. ItemMem attacks favor larger thresholds, with $\tau_m$ around $6$--$10$ for memorized items; and RecInertia attacks are most effective with the threshold $\tau_p$ between $0.6$ and $0.85$. We decide the parameter ranges with a validation set and then apply them to generate results in other figures. Notably, these threshold ranges remain largely stable across different models and prompt design variants, e.g., the number of shots and attack positions. Similarly, we include more details in the code repository.

We also examine the effects of parameter settings within their optimal ranges. By varying the decision parameters $\tau_m$ and $\tau_p$ within the valid ranges, we generate the log-scale ROC of these methods in Figure~\ref{fig:roc-auc-large}. We observe that both attacks exhibit remarkable performance, particularly in the low-false-positive area. 

\subsection{Attacks Under Defense}
We also experiment with existing defense methods for prompt extraction, focusing on prompt sanitization and output filtering, as mentioned earlier. For simplicity, we denote them as ``input'' and ``output'' defenses, accordingly. Figure~\ref{fig:prompt_defensed_combined} \& ~\ref{fig:agent_defensed_combined} presents the attack advantages after applying the defenses.

Since the similarity attack did not work, we first tested the input defense against the baseline inquiry attack. Figure \ref{fig:inquiry-defense} shows that inquiry attacks can be effectively defended with just input defense. 

We observe that input-based defenses perform more effectively on larger models. We attribute this to their stronger comprehension abilities, which enable them to better internalize and follow the defense instructions, such as avoiding repetition and generating more reasonable and diverse recommendations. Compared to smaller models, this enhanced understanding improves defense effectiveness. We posit that integrating a well-designed defense instruction tailored to a specific attack and dataset may constitute a pragmatic approach to mitigate privacy leakage. However, creating a universally applicable defense instruction requires further scrutiny and exploration.

The output defense works better in several cases (e.g., on the Beauty dataset), but struggles in others. We also found that large models perform better in output defense than small models in some cases. 

Overall, these results indicate that existing prompt-extraction defenses do not handle the unique structures in the ICL RecSys and thus have limited capabilities to defend against the proposed MIAs. 

\subsection{Factors Affecting Attacks}
\label{sec:factor_effect}
Section \ref{sec:attack-effectiveness} reports the best-performing result for each attack. In this section, we show the factors that affect ItemMem and RecInertia. Due to space limitations, we show only the results for the Movie dataset. We examine two factors: the number of shots and the location of the attacked example in the list of shots, which we believe are closely related to how the LLM responds to the attacks. We also investigate the number of poisoned items in the RecInertia attack. 

\definecolor{llama}{RGB}{31,119,180}
\definecolor{llama4}{RGB}{23,190,207}
\definecolor{gptoss20}{RGB}{214,39,40}
\definecolor{gemma}{RGB}{140,86,75}
\definecolor{mistral}{RGB}{0,0,0}
\definecolor{gptoss120}{RGB}{44,160,44}

\pgfplotsset{
    compact_axis/.style={
        width=0.90\linewidth,
        height=0.82\linewidth,
        scale only axis,
        ymin=0, ymax=1,
        tick label style={font=\tiny},
        label style={font=\tiny},
        xticklabel style={font=\tiny},
        yticklabel style={font=\tiny},
        legend style={font=\tiny},
        clip=true,
        enlarge x limits=false,
        enlarge y limits=false,
    },
    shots_axis_left/.style={
        compact_axis,
        xlabel={Number of Shots},
        ylabel={Attack Advantage},
        xmin=1, xmax=10,
        xtick={1,5,10},
        ytick={0,0.5,1.0},
    },
    shots_axis/.style={
        compact_axis,
        xlabel={Number of Shots},
        ylabel={},
        xmin=1, xmax=10,
        xtick={1,5,10},
        ytick={0,0.5,1.0},
    },
    position_axis_left/.style={
        compact_axis,
        xlabel={Position},
        ylabel={Attack Advantage},
        xmin=1, xmax=5,
        xtick={1,2,3,4,5},
        ytick={0,0.5,1.0},
    },
    position_axis/.style={
        compact_axis,
        xlabel={Position},
        ylabel={},
        xmin=1, xmax=5,
        xtick={1,2,3,4,5},
        ytick={0,0.5,1.0},
    },
    poison_axis_left/.style={
        compact_axis,
        xlabel={Poisoning Number},
        ylabel={Attack Advantage},
        xmin=1, xmax=10,
        xtick={1,5,10},
        ytick={0,0.5,1.0},
    },
    poison_axis/.style={
        compact_axis,
        xlabel={Poisoning Number},
        ylabel={},
        xmin=1, xmax=10,
        xtick={1,5,10},
        ytick={0,0.5,1.0},
    },
    llama/.style={color=llama, mark=o, thick},
    llama4/.style={color=llama4, mark=*, thick},
    gptoss20/.style={color=gptoss20, mark=square*, thick},
    gemma/.style={color=gemma, mark=triangle*, thick},
    mistral/.style={color=mistral, mark=diamond*, thick},
    gptoss120/.style={color=gptoss120, mark=+, thick},
}

\begin{figure*}[ht]
\centering

\begin{tikzpicture}
\begin{axis}[
    hide axis,
    width=\textwidth,
    height=0.15\textwidth,
    xmin=0, xmax=1,
    ymin=0, ymax=1,
    legend columns=3,
    legend style={
        font=\tiny,
        draw=none,
        at={(0.5,0.5)},
        anchor=center
    }
]
\addlegendimage{llama4}
\addlegendimage{mistral}
\addlegendimage{gptoss120}
\legend{Llama4, Mistral, GPT-OSS:120b}
\end{axis}
\end{tikzpicture}

\vspace{1mm}

\begin{minipage}[t]{0.18\textwidth}
\centering
\begin{tikzpicture}
\begin{axis}[shots_axis_left]
\addplot+[llama4]    coordinates {(1,0.63) (5,0.63) (10,0.45)};
\addplot+[mistral]   coordinates {(1,0.97) (5,0.99) (10,0.81)};
\addplot+[gptoss120] coordinates {(1,1.00) (5,0.99) (10,0.98)};
\end{axis}
\end{tikzpicture}

{\tiny (a) ItemMem}
\end{minipage}\hfill
\begin{minipage}[t]{0.18\textwidth}
\centering
\begin{tikzpicture}
\begin{axis}[shots_axis]
\addplot+[llama4]    coordinates {(1,0.54) (5,0.59) (10,0.42)};
\addplot+[mistral]   coordinates {(1,0.91) (5,0.89) (10,0.71)};
\addplot+[gptoss120] coordinates {(1,0.99) (5,0.96) (10,0.96)};
\end{axis}
\end{tikzpicture}

{\tiny (b) RecInertia}
\end{minipage}\hfill
\begin{minipage}[t]{0.18\textwidth}
\centering
\begin{tikzpicture}
\begin{axis}[position_axis]
\addplot+[llama4]    coordinates {(1,0.63) (2,0.68) (3,0.61) (4,0.67) (5,0.95)};
\addplot+[mistral]   coordinates {(1,0.95) (2,0.92) (3,0.88) (4,0.88) (5,0.97)};
\addplot+[gptoss120] coordinates {(1,1.00) (2,0.98) (3,0.99) (4,0.99) (5,0.99)};
\end{axis}
\end{tikzpicture}

{\tiny (c) ItemMem}
\end{minipage}\hfill
\begin{minipage}[t]{0.18\textwidth}
\centering
\begin{tikzpicture}
\begin{axis}[position_axis]
\addplot+[llama4]    coordinates {(1,0.59) (2,0.66) (3,0.61) (4,0.60) (5,0.92)};
\addplot+[mistral]   coordinates {(1,0.57) (2,0.59) (3,0.54) (4,0.57) (5,0.32)};
\addplot+[gptoss120] coordinates {(1,0.95) (2,0.92) (3,0.93) (4,0.97) (5,0.96)};
\end{axis}
\end{tikzpicture}

{\tiny (d) RecInertia}
\end{minipage}\hfill
\begin{minipage}[t]{0.18\textwidth}
\centering
\begin{tikzpicture}
\begin{axis}[poison_axis]
\addplot+[llama4]    coordinates {(1,0.54) (5,0.37) (10,0.06)};
\addplot+[mistral]   coordinates {(1,0.91) (5,0.92) (10,0.39)};
\addplot+[gptoss120] coordinates {(1,0.99) (5,0.52) (10,0.12)};
\end{axis}
\end{tikzpicture}

{\tiny (e) RecInertia}
\end{minipage}

\caption{Attack advantages under different prompt settings in recommendation systems: (a)–(b) impact of the number of shots on ItemMem and RecInertia, (c)–(d) impact of example position on ItemMem and RecInertia, and (e) impact of the poisoning number on RecInertia on the Movie dataset.}
\label{fig:prompt_five}
\vspace{-8pt}
\end{figure*}

\textbf{The Number of Shots}
We tested 1-shot, 5-shot, and 10-shot prompting strategies to evaluate model robustness. Figure~\ref{fig:prompt_five}(a) \& (b) illustrates the relationship between the number of shots in the prompt and the efficacy of ItemMem and RecInertia. The attack-targeted shot is put at the last position (the position effect will be studied next). Our analysis reveals that increasing the shots may have different effects on attacks and LLMs. ItemMem attacks remain consistently effective regardless of prompt length, indicating that memorized items can be reliably elicited even in the presence of expanded context. RecInertia attacks exhibit moderate sensitivity to the number of shots, a trend particularly pronounced in smaller models. These results demonstrate that prompt composition may play a critical role in affecting attack effectiveness. The observed decline in RecInertia's effectiveness can be attributed to increased informational load: as the context window expands with more demonstrations, the model's attention is distributed more broadly, thereby attenuating the impact of adversarial inputs. There is no clear pattern differentiating LLMs. 

\textbf{Effect of Attacked Position}
We also conduct attacks at each position within the 5-shot examples to examine their performance. Figure~\ref{fig:prompt_five} (c) \& (d) shows that the attack performance tends to vary more, either increasing or decreasing, around the last position.  Both ItemMem and RecInertia attacks maintain a stable performance until the last position. 

\textbf{Effect of Replaced Items}
The RecInertia attack perturbed the presented victim's interactions to see how the LLM responds. We investigate whether increasing the number of poisoned items affects the attack effectiveness. Figure \ref{fig:prompt_five} (e) shows that attack performance consistently decreases as more items are poisoned. This trend holds across all evaluated datasets and most models. 

\subsection{Member in Pretraining Data or Prompt?}
\label{sec:pretrained-RecSys}
A recent study \cite{10.1145/3726302.3730178} has shown that LLMs might have seen the popular RecSys datasets during pretraining. If these experimental datasets, Movie, Book, and Beauty, are in the pretraining dataset and the  LLM memorizes them very well, some of these MIA attacks might be affected. We looked into this issue with experiments. Specifically, the experiment can be described as follows. We provide the maximum amount of user context to check whether the LLM can recall the remaining part: if a user has $k$ interaction records, we show the LLM the first $k-1$ interactions and query whether it can correctly infer the $k$-th interaction. If the model produces the correct response, we consider it to have memorized the user and the associated interactions. The result indicates that the memorization effect of pretraining data is very weak. Specifically, on the ML-1M dataset, Llama:8b, Mistral and GPT-OSS:120b, demonstrate memorization rates of approximately 0.03\%, 0.18\% and 0.22\%, respectively, while no measurable memorization is observed on the Book and Beauty datasets. Therefore, we can conclude that our results on memorization-related attacks are less likely to be affected by the pretraining data memorization, and the membership inference on examples is dominated by the information in the prompt. 

\section{Related Work}
\subsection{MIA on RecSys}\label{sec:MIA-RecSys}
 The earlier RecSys MIA studies are focused on the user level. Zhang et al. \cite{zhang2021membership} propose the Item-Diff method for inferring membership in a target RecSys by analyzing the similarity between a user's historical interactions and recommended items. The core idea is that, for users in the training set, their historical interactions are likely to be more closely aligned with the items recommended by the system. Wang et al. 
 \cite{wang2022debiasing} propose the DL-MIA framework to improve Item-Diff with a VAE-based encoder and weight estimator to address issues with Item-Diff. More recently, Wei et al. \cite{yuan2023interaction} proposed a white-box interaction-level membership inference on federated RecSys.  Zhong et al. \cite{zhong2024interaction} proposed another interaction-level membership inference on Knowledge Graph-based RecSys, utilizing the similarity matrix between the interacted items and the recommended items. To our knowledge, no MIA study has been reported on LLM RecSys.

 \subsection{MIA on LLMs} \label{sec:MIA-LLM}
MIAs aim to determine whether a specific sample was included in a model's training data, posing significant privacy risks \cite{carlini2022lira, hu2022membership}. While widely studied in traditional machine learning \cite{hu2022membership,matsumoto2023,hayes2017,zhang2021membership}, MIAs have become increasingly concerning in large language models (LLMs), where revealing training data can leak sensitive information. Most MIA methods rely on the observation that models exhibit higher confidence on training samples \cite{carlini2022lira,hu2022membership}, leveraging access to output posteriors to compute signals such as loss \cite{duan2024privacyriskincontextlearning, wen2023standingcomparativeanalysissecurity} or perplexity \cite{274574,236216}. More advanced approaches incorporate intermediate representations \cite{Nasr_2019}, loss trajectories \cite{liu2022membershipinferenceattacksexploiting}, or shadow models \cite{carlini2022lira}, but generally require probability access. Recent work explores posterior-free MIAs \cite{Choquette-Choo21, li2021membershipleakagelabelonlyexposures} by estimating distances to decision boundaries, though these methods face challenges in black-box and discrete settings. Text-only MIAs have also been proposed for LLMs \cite{wen2024membershipinferenceattacksincontext}, but are limited to classification tasks and do not directly extend to recommendation scenarios due to the entirely different problem settings.

\section{Conclusion}\label{sec:conclusion}
ICL-based RecSys applies lightweight LLM customization, which has become an important research area due to its flexibility and low cost. However, its privacy risks have not been sufficiently studied. We designed novel MIA attacks, ItemMem and RecInertia, and demonstrated that they remain effective even when existing prompt-extraction defense methods are applied. The proposed attacks can be integrated into an agent-based attack-detection system to help defend against them. A more principled defense method, e.g., differentially private (DP) prompts \cite{tang2024privacypreserving}, might be employed to fundamentally address MIAs. However, a major concern is that DP-based methods will inevitably affect system utility. We will extend this study to include more MIA attacks and better mitigation methods. 


\bibliographystyle{ACM-Reference-Format}
\bibliography{custom}

\end{document}